# Geometry-aware neural causal discovery for large-scale spatiotemporal systems

Haoyang Yan[1,2], Kaiqi Zhao[1], Weiping Wang[3,*], Yunpeng Wang[1,4*], Xiaolei Ma[1,5*]

[1] *School of Transportation Science and Engineering, Beihang University, Beijing 100191, China*

[2] *Key Laboratory of Spectral Imaging Technology, Xi'an Institute of Optics and Precision Mechanics, Chinese Academy of Sciences, Xi' an, China*

[3] *The National Key Laboratory of Autonomous Intelligent Unmanned Systems, Tongji University, Shanghai 200120, China*

[4] *State Key Laboratory of Intelligent Transportation System, Beijing 100191, China*

[5] *Key Laboratory of Intelligent Transportation Technology and System, Ministry of Education, Beijing 100191, China*

* Corresponding author. Email: weipwang@tongji.edu.cn; ypwang@buaa.edu.cn; xiaolei@buaa.edu.cn

## Abstract

Causal discovery at large spatiotemporal scale is difficult: variables are physically embedded, candidate interactions grow quadratically with system size, and causal structure changes with system state. We introduce GeoDCD, a geometry-aware neural framework that uses spatial coordinates to initialize a learnable hierarchy and converts a trained nonlinear predictor into time-varying directed graphs through input-output Jacobian sensitivity analysis. On chaotic Lorenz-96 dynamics, GeoDCD attains an F1 score of 0.99, reducing structural Hamming distance by 36.8% relative to the strongest neural baseline, and still leads flat baselines when coordinates are uninformative. Runtime scales approximately linearly over the evaluated range, enabling discovery on a 10,512-node sea-level-pressure grid. Applied to observations, GeoDCD identifies circulation-consistent gateways, resolves El Niño/La Niña-dependent reorganization, and separates energy-to-traffic from traffic-to-energy influence in coupled electric-vehicle and road systems. Edges are neural-Granger sensitivities rather than interventional effects, positioning GeoDCD for mechanistic hypothesis generation where interventions are unavailable.

## Introduction

Understanding how complex systems evolve and respond to perturbations requires uncovering not only statistical associations but also the underlying causal mechanisms governing interactions among system components[1]. This need is especially pressing in large-scale spatiotemporal systems, where physically embedded variables interact through nonlinear feedback, multi-scale dependencies and state-dependent causal structures[2,3]. As machine learning increasingly drives scientific discovery and decision-making, causal inference in large-scale spatiotemporal systems—from climate dynamics[4] and ecological networks to gene regulation[5], epidemiology[6] and socio-technical infrastructures—has become central to building reliable and actionable machine intelligence.

Recent advances in causal representation learning, deep causal discovery and stable learning have emphasized the convergence of causal inference and machine learning[7], and the role of causal reasoning in actionable prediction and dataset-bias control[8]. Together these define a target for scientific machine intelligence: causal discovery should be nonlinear, scalable and interpretable, while remaining explicit about the assumptions under which directed structure can be recovered. Classical methods, including Granger causality and structural equation modelling, rely on linearity, stationarity or limited dimensionality. Convergent cross mapping detects nonlinear causal interactions but is sensitive to synchrony, indirect effects and observational noise[9]; partial cross mapping and model-based frameworks[10] improve direct-cause identification by conditioning on mediators[11,12], but remain constrained by long time-series requirements and limited scalability.

These limitations are acute at large scale. Candidate directed interactions grow quadratically with the number of variables, making exhaustive conditional testing or unconstrained neural graph learning computationally expensive and statistically unstable. Although deep learning helps under data limitations[5], flat graph learning is poorly matched to systems in which variables occupy metric space and interactions are organized across local, regional and global scales. Geometry is not merely metadata: it encodes proximity, transport pathways, functional modularity and hierarchical organization, providing structural priors that reduce the effective search space of causal discovery. Recent geometric deep learning studies demonstrate the value of such priors for microscopic motion[13] and urban road structure[14].

A second gap is temporal variability. Causal mechanisms are often not fixed: interactions may strengthen, weaken or reverse across regimes, and feedback pathways may emerge only under specific states. Reviews of time-series causal inference stress that temporal ordering, conditioning, confounding and stationarity assumptions must be handled explicitly[12]. In Earth system science, causal networks have identified gateways propagating sea surface temperature anomalies across ocean basins[11], evaluated whether climate models reproduce observed interaction pathways[15], and inferred spatial causal links from geographical information[16]. Yet most approaches target small or medium systems, static graphs or domain-specific model classes, and even high-dimensional neural causal learning[5] remains insufficiently adapted to large spatiotemporal systems where geometry and state-dependent causal structure are both central.

Here we introduce GeoDCD, a geometry-aware neural causal discovery framework for large-scale spatiotemporal systems. Spatial coordinates organize variables into multi-resolution patches; a causal Transformer learns temporal dependencies; sparse basis decomposition represents dynamic graph structure; and Jacobian-based causal effect extraction converts learned dynamics into directed, time-varying causal graphs. This replaces unconstrained dense graph learning with multi-resolution causal structure learning, enabling recovery of instantaneous and state-dependent causal interactions among thousands of variables. We evaluate GeoDCD on controlled Lorenz-96, Cluster-Lorenz and Finance benchmarks and on real-world climate and energy–mobility systems[17–19], which serve as stress tests for scale, non-stationarity and transfer rather than as isolated applications.

## Framework overview

GeoDCD learns dynamic causal networks from two inputs: a multivariate sequence with T time steps and N variables, and a coordinate matrix giving the location or embedding of each variable — latitude–longitude positions, road-sensor locations, or virtual coordinates when no physical embedding is available. It infers both a persistent N-by-N directed causal graph and a time-resolved N-by-N causal effect tensor (Fig. 1). Mathematical details, training losses and implementation settings are given in Methods.

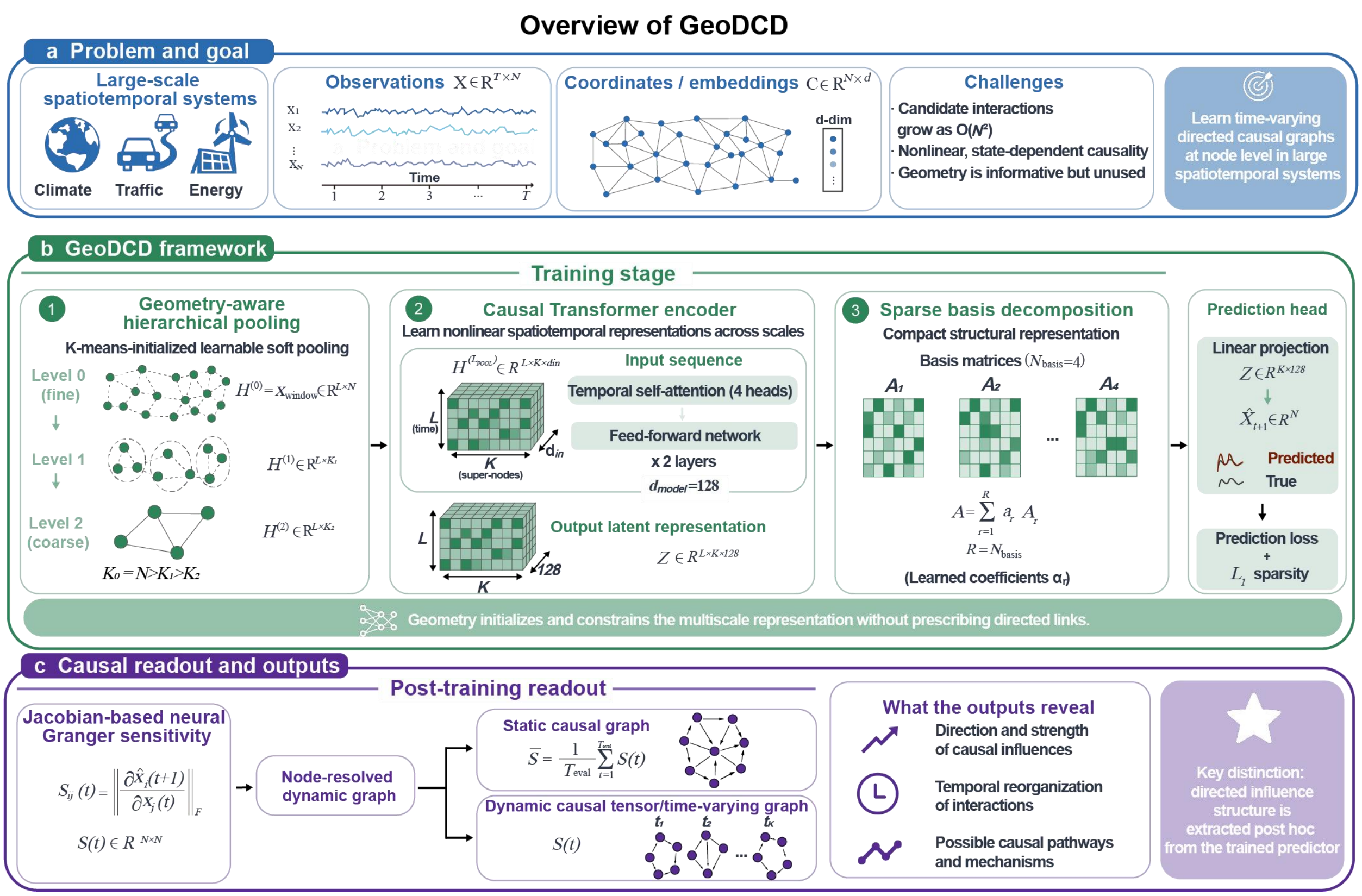

*Fig. 1 | Overview of GeoDCD. a, Problem formulation. GeoDCD takes multivariate observations X ∈R and node coordinates or embeddings C ∈R to recover node-resolved, time-varying directed causal graphs in large spatiotemporal systems. b, Training framework. K-means-initialized learnable geometric pooling constructs multiscale representations, a causal Transformer learns nonlinear temporal dependencies, and sparse basis decomposition parameterizes the interaction structure before next-step prediction. Geometry initializes and regularizes the hierarchy but does not prescribe directed links. c, Post-training causal readout. Jacobian-based sensitivity analysis projects learned effects back to individual nodes, producing an aggregated causal graph G ∈R and a dynamic causal tensor S(t), which characterize causal direction, strength and temporal reorganization.*

The causal discovery problem is formulated in the neural Granger-causality paradigm. A directed edge $j \rightarrow i$ is identified when the history or current state of node j contributes to the learned prediction of the future state of node $i$ after conditioning on the multivariate temporal context. This framing is important for large observational systems, where controlled interventions are usually unavailable and where physical mechanisms may be nonlinear, non-stationary and distributed across multiple scales. GeoDCD therefore does not seek a static correlation network. It learns a predictive dynamical model whose local input-output derivatives define directed causal effects, allowing causal structure to change with system state. The resulting input – output sensitivity is interpreted as a state-dependent neural Granger score within the fitted predictive system.

Geometry enters as a causal-discovery prior rather than a constraint. Standard neural graph learning treats all node pairs as candidate interactions, creating an $O(N^2)$ search space that encourages dense spurious links; GeoDCD instead normalizes coordinates and partitions nodes into a multi-resolution hierarchy of spatial patches, so that local, regional and long-range effects are represented at different scales (Supplementary Figure 1). A causal Transformer with masked self-attention then maps sliding windows to latent dynamics, replacing the linear autoregressive map of classical Granger causality with a nonlinear sequence model, and a sparse basis decomposition parameterizes the effective graph layer using shared adjacency bases rather than one dense weight per node pair. Training combines next-step prediction loss with an L1 sparsity penalty, and Jacobian sensitivity analysis of the trained predictor converts learned dynamics into a directed dynamic adjacency tensor, aggregated across time for static structure and retained per window for transient changes. Crucially, the hierarchy is an inductive bias, not a hard causal constraint: when geometry is weak or artificial, the temporal model and sparse structural learning can still recover intrinsic topology from the data, so the causal graph is ultimately learned from observed dynamics rather than prescribed by distance alone.

The geometric hierarchy enables GeoDCD to scale to large spatial systems by organizing variables into geometry-aware clusters and learning dynamic dependencies across multiple resolutions, rather than evaluating all pairwise candidate interactions in a flat dense graph. This K-means-initialized, trainable organization is illustrated by San Diego with 716 nodes (**Fig.2**a), Beijing with 513 nodes (Fig. 2b), Greater Los Angeles with 3,834 nodes (Fig. 2c), and the Bay Area with 2,352 nodes (Fig. 2d). The resulting partitions reflect compact urban regions, transport corridors and branching structures. Because the input – output Jacobian is taken with respect to the original node-level input histories and node-level predictions, it directly yields node-resolved sensitivities while propagating gradients through all multiresolution components of the trained predictor. Applying this readout to the San Diego and Beijing networks yields cluster-ordered causal intensity matrices in which inferred causal strength concentrates within geometric partitions and reorganises across consecutive inference windows (Supplementary Fig. 2), confirming that the partitions in Fig. 2a-d are not merely a computational device but coincide with the structure of the recovered causal graph. Accordingly, mean training time per epoch ranges from 24.2 s in Beijing and 65.0 s in San Diego to 242.0 s in the Bay Area and 366.4 s in Greater Los Angeles (Fig. 2e), following an approximately linear relationship with network size ($R^2$=0.991); the largest network requires about 10.2 h for 100 epochs. These results indicate that geometric hierarchy is not an auxiliary preprocessing step, but a core architectural component that makes large-scale dynamic causal discovery computationally practical.

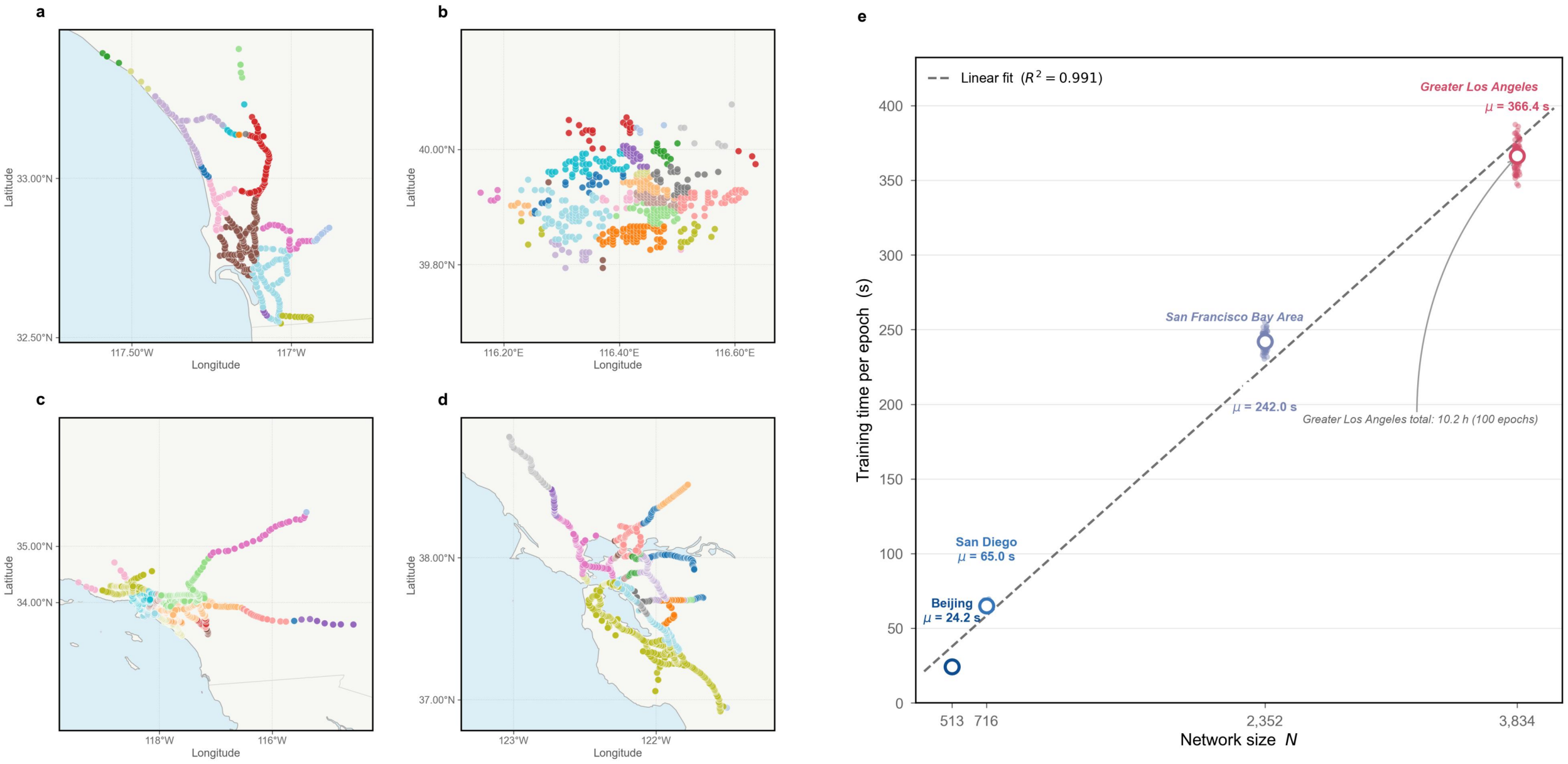

*Fig. 2 | Geometry-aware hierarchical aggregation and computational scaling across urban road networks. a–d, Coordinate-based node aggregation inferred by GeoDCD for four road-network systems: San Diego with 716 nodes (a), Beijing with 513 nodes (b), Greater Los Angeles with 3,834 nodes (c), and the San Francisco Bay Area with 2,352 nodes (d). Nodes are plotted at their geographical coordinates, and colours denote the inferred cluster assignments. The resulting partitions follow spatially coherent road-network structures, including compact urban regions, branching transport corridors and large ring-road-like organization, while reducing networks containing hundreds or thousands of nodes to a tractable set of geometric groups. e, Training time per epoch as a function of network size. Small points denote epoch-level runtimes and open circles indicate the mean for each network. Mean training time increases from 24.2 s for Beijing and 65.0 s for San Diego to 242.0 s for the San Francisco Bay Area and 366.4 s for Greater Los Angeles. The dashed line shows the least-squares linear fit across all runtime samples ( $R^2$=0.991); training the largest system for 100 epochs requires approximately 10.2 h. The observed relationship supports near-linear execution scaling over the evaluated range, in contrast to the quadratic candidate-pair growth associated with flat all-pairs formulations. The fitted trend describes GeoDCD runtime only; quadratic neural baselines are not plotted.*

# Results

## Simulation benchmarks

We first evaluated GeoDCD where the true directed graph is known, so the output can be assessed as causal structure rather than forecasting skill. The suite spans four regimes: VAR provides a linear baseline with known coefficient matrices; Lorenz-96 introduces nonlinear chaotic feedback on a circular manifold; Cluster-Lorenz creates localized patches with dense within-cluster and sparse between-cluster coupling; and Finance removes meaningful physical geometry by assigning random virtual coordinates, testing whether the geometric prior becomes harmful when the coordinate signal is uninformative.

Following the simulation-benchmark logic used in high-dimensional neural causal discovery studies, each method was trained and evaluated against the ground-truth adjacency structure. We compared GeoDCD with linear Granger-style models and nonlinear neural causal discovery baselines, including GVAR, cMLP, cLSTM, TCDF, eSRU, NAVAR variants, CUTS+, UnCLE and Causalformer. Performance was summarized with AUROC for edge discrimination, F1-score for binary causal link recovery and Structural Hamming Distance (SHD) for the number of edge additions, deletions or reversals required to transform the learned graph into the true graph. Results were averaged over five independent random seeds using the same preprocessing and evaluation protocol. Full numerical results for all ten baselines and GeoDCD across the four benchmarks are reported in Supplementary Table 1.

The clearest test is Lorenz-96, where causal links arise from nonlinear chaotic dynamics and the true graph follows the circular coupling structure. Linear Granger modelling fails to recover this topology (GVAR F1 = 0.42), confirming that the task is not reducible to a linear autoregressive coefficient test. Neural baselines improve by learning nonlinear dynamics, but without geometric organization they still introduce many spurious edges. GeoDCD recovers the Lorenz-96 topology with F1 = 0.99 and SHD = 9.6, reducing SHD by 36.8% relative to the strongest comparator, Causalformer (SHD = 15.2). This result supports the central claim that coordinate-space hierarchy constrains the nonlinear causal search space rather than merely improving prediction as shown in **Fig. 3**.

Cluster-Lorenz tests a different failure mode: causal structure is not only nonlinear but multi-scale. The ground truth contains dense local causal interactions inside spatial patches and sparse longer-range interactions between patches. Flat neural models must infer this grouping implicitly from time series alone, which tends to mix local coupling with cross-cluster influence. GeoDCD's hierarchical geometric pooling makes the patch structure explicit before causal graph learning, yielding F1 = 0.95. The result indicates that the geometry module improves direct edge recovery in systems where causal mechanisms are organized by local neighbourhoods and regional modules.

Finally, the Finance benchmark stress-tests the opposite condition. Variables have no true physical layout, and the assigned coordinates are random virtual labels rather than meaningful geometric information. If GeoDCD simply overfit spatial proximity, performance should degrade sharply in this setting. Instead, GeoDCD retains F1 = 0.78 and AUROC = 0.96, outperforming flat neural baselines such as Causalformer (F1 = 0.72) and CUTS+ (F1 = 0.70). The controlled benchmarks therefore separate three claims: GeoDCD can recover nonlinear directed topology, exploit multi-scale geometric organization when it is informative and avoid collapse when geometry is deliberately uninformative.

To separate the contribution of the geometric module from that of the multi-scale hierarchy, we ablated each independently. Replacing the learnable coordinate-driven pooler with a fixed, coordinate-agnostic partition while retaining the hierarchy ("No-Geometric") reduced F1 on Lorenz-96 from 0.99 to 0.77 and increased SHD from 9.6 to 202. Removing the hierarchy while retaining learned geometric clustering ("Flat") left F1 at 0.96 and SHD at 31. Geometry, rather than hierarchical composition per se, is therefore the dominant source of recovery accuracy (Supplementary Note 1 and Supplementary Table 2).

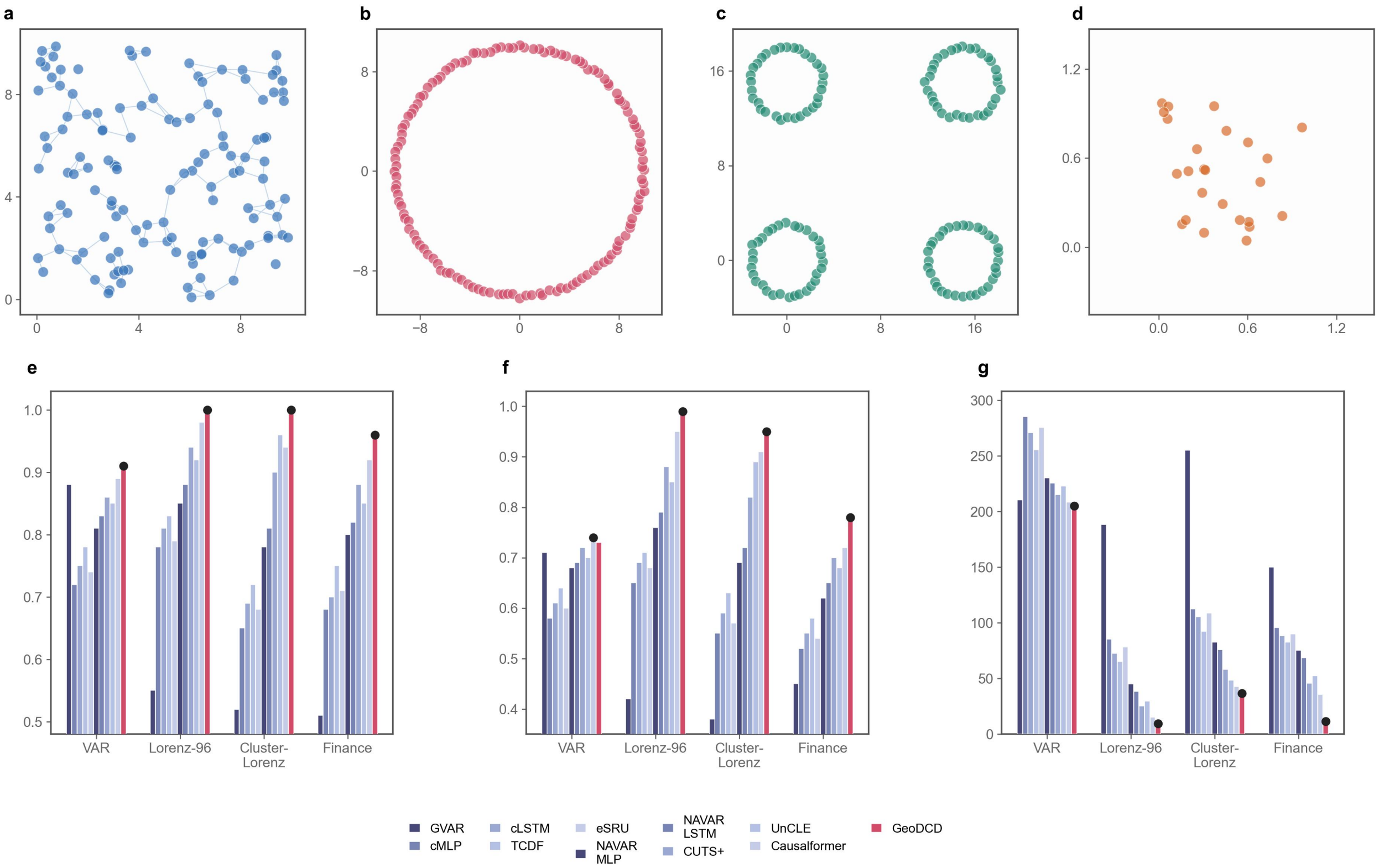

*Fig. 3 | Causal graph recovery on controlled benchmarks. a–d, Geometric organization of the four benchmark systems. a, Spatially embedded vector autoregressive (VAR) system with (N=128) variables on a uniform grid; lines denote ground-truth graph connections and circles denote individual variables. b, Lorenz-96 system with (N=128) variables arranged on a circular manifold. c, Cluster-Lorenz system comprising (N=128) variables distributed across four spatially separated ring clusters. d, Finance system with (N=25) variables and no prescribed spatial prior. Colours distinguish the four geometric configurations and do not encode performance. e–g, Causal graph recovery performance of GeoDCD and ten baseline methods across the VAR, Lorenz-96, Cluster-Lorenz and Finance benchmarks. e, Area under the receiver operating characteristic curve (AUROC). f, F1 score. g, Structural Hamming distance (SHD). Higher AUROC and F1 scores indicate more accurate recovery, whereas lower SHD indicates fewer graph-editing errors. GeoDCD is shown in rose and baseline methods in blue shades; black circles identify the best-performing method for each benchmark. Across the four systems, GeoDCD obtains AUROC values of 0.91, 1.00, 1.00 and 0.96; F1 scores of 0.73, 0.99, 0.95 and 0.78; and SHD values of 205.2, 9.6, 36.8 and 11.5, respectively. Bars report means across five independent random seeds. Numerical values are listed in Supplementary Table 1.*

## Large-scale system evaluation for global climate data

We next evaluated whether GeoDCD scales to a global physical system whose causal structure is high-dimensional, non-stationary and only partially verifiable through domain knowledge. No complete ground-truth causal graph exists for the atmosphere, so the relevant test is whether the inferred network remains computationally feasible, physically interpretable and consistent with established process knowledge — the same logic used in process-oriented climate model evaluation and time-series causal inference in Earth systems[15].

We applied GeoDCD to NCEP/NCAR reanalysis sea-level pressure (SLP) data on a 2.5° global grid, yielding a network with 10,512 spatial nodes. At this scale, a flat dynamic causal discovery model would require evaluating an enormous set of candidate directed interactions and would be prone to dense, incoherent links. GeoDCD instead uses coordinate-aware pooling to organize the grid into local and regional representations before extracting time-varying causal effects. The climate experiment therefore serves as the primary large-scale system test: it evaluates whether the same neural causal discovery mechanism that recovers synthetic topology can operate on a planetary graph.

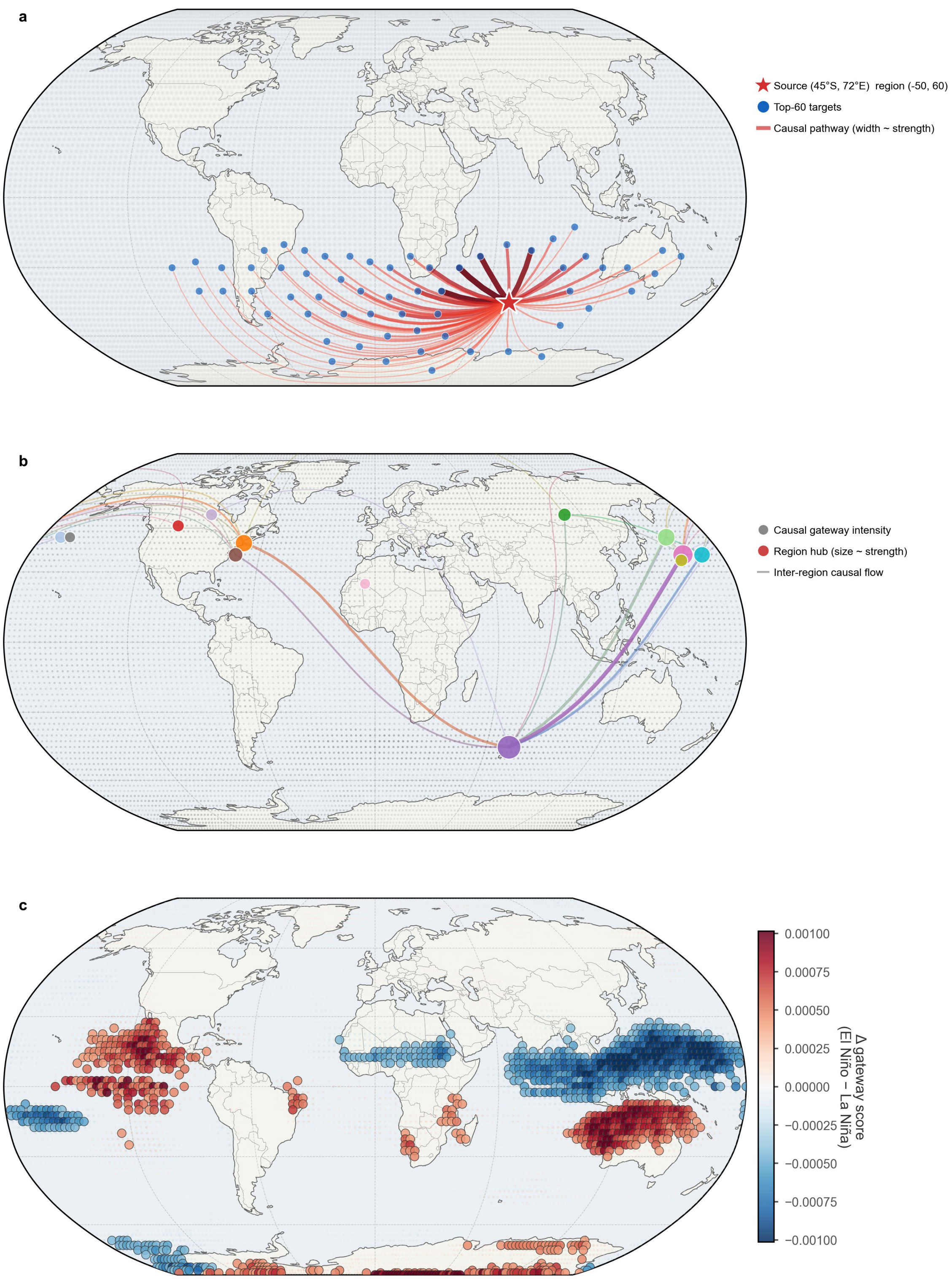

*Fig. 4 | Multiscale dynamic causal discovery in global sea-level pressure. a, Directed causal-flow pathways recovered by GeoDCD from the NCEP sea-level-pressure field comprising 10,512 spatial grid points. The red star marks the primary causal gateway at 45$^0$ S, 72$^0$ E, selected from the leading gateway region, and blue circles denote 60 geographically separated, remotely reachable target nodes. Curved great-circle connections represent directed pathways retained from the strongest 10% of graph edges; line width and colour intensity increase with pathway strength. b, Hierarchical organization of the inferred global causal graph. Grey background points show node-level causal gateway intensity, defined from weighted outgoing connectivity. Coloured circles identify the representative hubs of 15 level-1 macroregions, with marker size proportional to combined incoming and outgoing causal strength. Curved links show the strongest 30% of causal interactions aggregated between macroregions, illustrating how node-level dependencies are compressed into a geographically interpretable multiresolution network. c, Reorganization of causal gateway topology between El Niño and La Niña winters. Colours represent the difference in mean November–February gateway scores between El Niño and La Niña composites. Each composite averages 32 monthly fields from eight event years: 1972, 1982, 1987, 1991, 1997, 2002, 2009 and 2015 for El Niño, and 1973, 1975, 1988, 1998, 1999, 2007, 2010 and 2020 for La Niña. Red values indicate stronger gateway activity during El Niño, whereas blue values indicate stronger activity during La Niña. Large outlined markers highlight grid points in the upper decile of absolute phase-dependent change; smaller translucent points provide the global background distribution.*

**Fig. 4**a shows the first large-scale finding: the emergence of interpretable global causal gateways without climatological supervision. The strongest inferred gateway appears near 45°S, 72°E in the Southern Indian Ocean. From this source, directed arcs propagate eastward and northward, following pathways consistent with the Roaring Forties, the Antarctic Circumpolar Current and atmospheric wave-guide dynamics. This is a plausible pattern because localized Southern Ocean dynamics can have global influence, and amplified Rossby-wave patterns provide a known mechanism for coherent remote atmospheric responses[20,21]. Thus, GeoDCD does not produce a dense global association map; it concentrates causal outflow into sparse routes linking pressure perturbations to downstream regions including Western Australia, South Africa and Patagonia.

**Fig. 4**b evaluates whether the geometric hierarchy remains meaningful at planetary scale. After coordinate-aware pooling, the 10,512-node grid is compressed into spatial units that preserve recognizable atmospheric organization rather than arbitrary coordinate clusters. The resulting regions reconstruct zonal pressure structures, including subtropical high-pressure ridges and subpolar low-pressure belts. Their annular organization is consistent with Southern Hemisphere circulation variability, particularly the Southern Annular Mode, while the concentration of inter-cluster flow over Southern Ocean and continental regions is compatible with the global reach of Southern Ocean dynamics[21,22]. This subfigure therefore links the method's scalability mechanism to interpretable climate structure.

**Fig. 4**c addresses the dynamic part of the evaluation. We compared inferred causal graphs during El Niño and La Niña periods and mapped the difference in causal gateway strength. During El Niño, the tropical Pacific becomes a stronger causal driver, with eastward pathways transmitting SLP anomalies toward the Americas. During La Niña, this equatorial dominance weakens and the graph shifts toward stronger subtropical coupling. The direction of this contrast is consistent with ENSO's role as a leading mode of interannual variability and with recent syntheses describing its spatially complex and state-dependent teleconnections[23,24].

## Large-scale cross-modal systems evaluation for energy-mobility infrastructure

We finally evaluated GeoDCD on Shenzhen's coupled energy-mobility infrastructure, focusing on the cross-modal EV-traffic causal graph rather than on single-domain traffic transfer. The ST-EVCDP cross-modal charging–traffic data were represented as a 494-node dynamic graph by pairing road-traffic occupancy and electric-vehicle charging energy volume across 247 traffic analysis zones. Each zone therefore contributes two interacting modalities embedded in the same urban coordinate system. This setting is a stronger test than ordinary traffic forecasting: a learned edge can connect traffic to traffic, energy to energy, traffic to energy or energy to traffic, and the model must distinguish these directions from the same observational sequence.

This evaluation is motivated by the tightening coupling between travel demand, charging infrastructure and electricity operations: EV usage and charging demand are heterogeneous across vehicle types, cities and activity patterns, and spatially concentrated demand can materially alter grid stress[25–27]. Charging demand should therefore not be treated as an isolated energy time series, nor congestion as an isolated mobility signal.

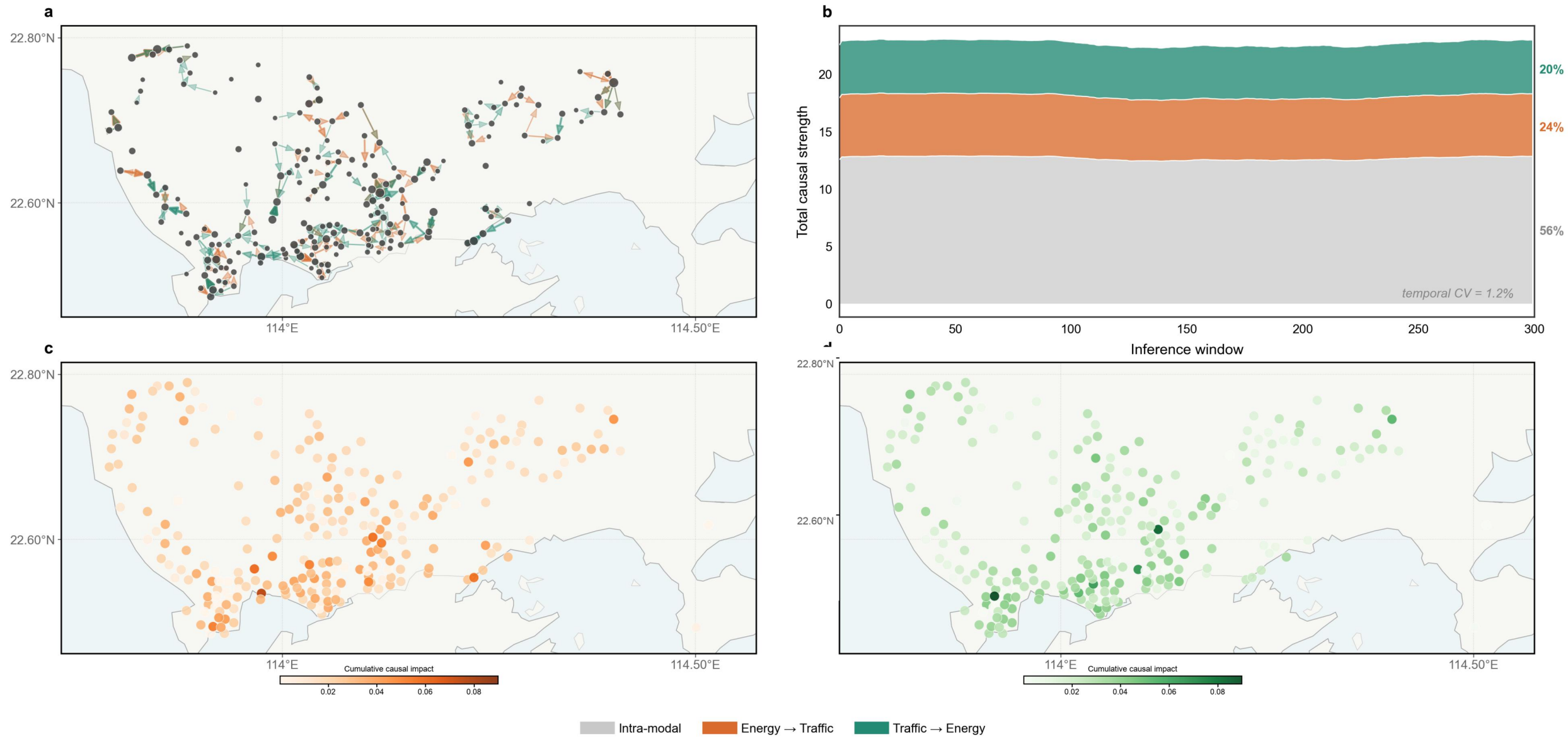

*Fig. 5 | Cross-modal EV–traffic causal graph in Shenzhen. a, Spatial organization of the strongest directed causal links between road-traffic occupancy and EV charging energy volume across 247 traffic analysis zones in Shenzhen. The 180 strongest links retained in each direction after 94th-percentile thresholding are shown, with orange arrows denoting energy-to-traffic effects and teal arrows denoting traffic-to-energy effects. Arrow width and opacity scale with inferred causal strength, whereas node size reflects the total strength of incident cross-modal connections. b, Temporal evolution of causal-strength allocation across 300 inference windows. Intra-modal dependencies account for approximately 56% of the total causal budget, compared with 24% for energy-to-traffic coupling and 20% for traffic-to-energy coupling. The low temporal coefficient of variation (1.2%) indicates that the aggregate directional composition remains stable despite local temporal fluctuations. c, Spatial distribution of cumulative energy-to-traffic causal influence, revealing geographically heterogeneous concentrations of charging-related effects on traffic conditions. d, Corresponding distribution of cumulative traffic-to-energy influence, showing locations at which traffic dynamics exert the strongest aggregate effects on charging demand. Panels c and d use a common colour scale to facilitate direct comparison of the two causal directions.*

**Fig. 5**a shows the inferred cross-modal causal graph over the Shenzhen road network. GeoDCD does not collapse the two modalities into an undirected association map. Instead, it distinguishes causal edges from traffic to energy and from energy to traffic while preserving the geographic embedding of traffic analysis zones. The strongest links are sparse and spatially structured, indicating that the model concentrates cross-modal influence in a limited set of infrastructure corridors and charging-demand zones rather than assigning diffuse influence across all co-located variables.

The causal coupling budget in **Fig. 5**b shows a remarkably stable allocation across 300 consecutive inference windows: intra-modal coupling consistently absorbs $\sim 56\%$ of total causal attention, while the remaining $\sim 44\%$ splits asymmetrically between energy-to-traffic ($\sim 24\%$) and traffic-to-energy ($\sim 20\%$) directions. The low temporal coefficient of variation (1.2%) indicates that this structural coupling is an intrinsic property of the infrastructure topology rather than a transient regime. The persistent directional asymmetry—energy fluctuations exert slightly stronger downstream effects on traffic than vice versa—reveals a functional hierarchy that static correlation analyses cannot disentangle.

Figures 5c and 5d reveal a spatially asymmetric bidirectional structure. Energy-to-traffic influence concentrates in the high-density western urban core, including Nanshan and Futian, where charging-load fluctuations can coincide with local traffic anomalies near dense activity centres. Traffic-to-energy influence is stronger along eastern corridors, including Luohu and Longgang, where congestion can redirect downstream charging behaviour. This asymmetry is consistent with the view that EV charging demand is spatially heterogeneous and infrastructure-dependent[25], while grid impacts depend strongly on where charging demand accumulates[27].

These results are not intended as a separate domain-specific contribution. They demonstrate that the same geometry-aware neural causal discovery mechanism transfers from controlled synthetic dynamics to climate and energy–mobility systems without redesigning the model for each application, and that for coupled infrastructure GeoDCD identifies not only where charging demand and traffic flow are associated, but which modality is more likely to drive the other.

## Discussion

This study establishes GeoDCD as a geometry-aware framework for dynamic causal discovery in large spatiotemporal systems. The central result is not that a neural model can predict future observations, but that the structure used to make those predictions can be converted into sparse, directed and time-resolved causal graphs. This addresses a gap between the use of deep learning for causal structure learning under limited data[5] and the need for causal reasoning in reliable scientific and operational decision-making[28].

The empirical findings show why geometry is useful for causal discovery rather than merely for representation learning. In Lorenz-96, GeoDCD recovers the nonlinear circular topology with F1 = 0.99 and reduces SHD relative to the strongest comparator, indicating that the geometric prior constrains the search space without replacing temporal evidence. In Cluster-Lorenz, the same mechanism separates dense local coupling from sparse cross-cluster influence. In Finance, where coordinates are deliberately uninformative, performance decreases but does not collapse, showing that GeoDCD does not simply equate spatial proximity with causality. This pattern supports the interpretation that geometry acts as an inductive bias: it regularizes plausible interactions when the embedding is meaningful, while the learned temporal dynamics remain responsible for edge discovery.

The real-system analyses extend this argument from controlled topology recovery to interpretable scientific structure. In global sea-level pressure data, GeoDCD processes a 10,512-node grid and identifies sparse causal gateways, hierarchical atmospheric regions and ENSO-dependent graph reorganization. These findings are consistent with the broader view that Earth-system dynamics are organized by nonlinear, state-dependent couplings rather than isolated variables[12,17]. Recent work on coupled climate modes similarly argues that predictability in the climate system can emerge from learning interactions among modes rather than modelling each component separately[4]. GeoDCD is aligned with this systems view, but shifts the target from forecasting coupled variables to recovering directed dynamic causal pathways among spatial nodes.

The Shenzhen experiment illustrates the value of causal graphs for coupled infrastructure. EV charging demand and traffic occupancy are not independent urban signals: charging behaviour depends on mobility patterns, while charging activity can create localized access traffic and infrastructure load. The inferred graph distinguishes traffic-to-energy from energy-to-traffic links, revealing a western pattern dominated by energy-driven traffic influence and an eastern pattern dominated by traffic-driven energy demand, consistent with evidence that charging demand is spatially heterogeneous[26], that infrastructure operation affects grid impacts[25] and that concentrated demand contributes to distribution-grid congestion[27]. GeoDCD does not replace transportation or power-system modelling; it provides a cross-modal causal layer that can expose when one subsystem appears to drive another.

This positioning distinguishes GeoDCD from black-box spatiotemporal forecasting models. Data-driven weather models achieve remarkable predictive skill at operational scales[18,29], but forecast accuracy does not by itself reveal the mechanisms that make the prediction possible. GeoDCD treats prediction accuracy as an intermediate objective: predictable variance becomes the route through which directed influence is estimated, in line with a broader agenda in which inductive biases and physical structure make learned models more interpretable and mechanistically informative[17,30]. The Jacobian graph therefore provides a dynamic readout of the learned nonlinear system, enabling inspection of transient propagation, regime-dependent teleconnections and cross-modal feedback.

Several limitations define the scope of the claims. First, GeoDCD operates within a neural Granger-causal setting: an edge is inferred when a source variable sensitively affects the learned prediction of a target after conditioning on the observed multivariate context. This is a pragmatic observational definition of causal influence, not a substitute for intervention; hidden common drivers, omitted variables, synchrony and temporal aggregation can affect graph recovery[31]. Second, the usefulness of hierarchical geometric pooling depends on whether the coordinate system is scientifically meaningful — climate grids and traffic analysis zones provide natural embeddings, but domains where functional connectivity diverges from physical distance may require learned or hybrid geometries, and simple coordinate clustering can produce regions that are geometrically compact but dynamically incomplete. Third, the number of adjacency bases trades expressivity against sparsity. Finally, full Jacobian extraction remains expensive at very large scale, and differences in sampling rate, noise level and measurement scale across modalities can bias causal strengths unless normalization and modality-aware weighting are handled carefully.

Future work should incorporate uncertainty estimates for dynamic edges, stronger controls for hidden confounding, continuous-time extensions for irregularly sampled systems, and online adaptation so that inferred graphs update under distribution shift without complete retraining. Testing on further coupled domains, including power-grid stability, ecological food webs and neuronal circuits, would establish how far geometry-aware causal discovery generalizes.

# Methods

## Causal discovery formulation

The Framework overview describes the conceptual role of geometry, temporal encoding and differentiable causal readout. Here we specify the operational formulation used in the experiments. We consider a multivariate spatiotemporal system with an observation matrix $D \in R^{T\times N}$, where T is the number of time steps and N is the number of variables or spatial nodes. At time t, the observed state is $x_t = D_{t,:} \in R^N$. Each node has an associated coordinate vector, collected in $C \in R^{N\times d}$; $d = 2$ for latitude-longitude or planar sensor coordinates and may be larger when an application supplies a richer embedding.

GeoDCD follows a neural Granger-causality formulation. In a linear vector autoregressive model, causal structure is encoded by lagged coefficient matrices: a source variable j does not Granger-cause target i if all corresponding lag coefficients vanish. GeoDCD replaces the linear autoregressive map with a nonlinear predictor $f_\theta$ while preserving the same predictive principle: an edge j → i is supported when variation in the history or current state of node j affects the modelled future state of node i after conditioning on the multivariate context.

$$x_t = \sum_{\tau=1}^{P} A_\tau x_{t-\tau} + \varepsilon_t$$

$$\hat{x}_{t+1} = f_\theta(x_t, x_{t-1}, \dots, x_{t-P+1})$$

The target of inference has two forms. The static graph $A \in R^{N\times N}$ summarizes persistent source-target relations aggregated across time. The dynamic causal tensor $S(t) \in R^{N\times N}$ records instantaneous causal strength at each inference time. The experiments therefore evaluate both edge recovery in settings with known ground truth and dynamic graph interpretability in real systems where complete causal labels are unavailable.

$$S_{i\leftarrow j}(t) = \| \frac{\partial \hat{x}_{i,t+1}}{\partial x_{j,t}} \|_F$$

## Input tensor construction and normalization

The raw observation matrix is transformed into overlapping temporal windows before training. For a look-back length L, each training item contains an input tensor $X_b \in R^{N\times L}$ and a one-step-ahead target $y_b \in R^N$. This conversion yields a batch tensor $X \in R^{B\times N\times L}$, where B is the number of windows. For traffic and other regularly sampled systems, windows are created chronologically to avoid future leakage; the traffic experiments use $L = 12$ and stride 6. The model is trained to predict the next observation from the current window, so causal extraction is tied to the same temporal context used for prediction.

All time series are standardized using statistics computed on the training split. Coordinates are also standardized before clustering so that spatial partitioning is not dominated by units of measurement or map scale. For climate grids, latitude-longitude coordinates define the embedding. For road networks and traffic analysis zones, projected sensor or zone coordinates are used. In non-spatial robustness tests such as the Finance benchmark, random virtual coordinates are supplied deliberately to test whether the model can avoid over-reliance on uninformative geometry.

## Hierarchical geometric pooling

Hierarchical geometric pooling is implemented as K-means-initialized learnable soft clustering, not as an additional learned causal claim. At hierarchy level $l$, the normalized node coordinates $C^{(l)}$ are first partitioned by K-means to initialize $K_{l+1}$ trainable centroids. K-means is applied only for initialization and does not fix the assignments used by the trained model. During training, node-to-patch assignments are refined jointly with the predictive model through the centroid locations, a coordinate-conditioned refinement network and a learnable temperature parameter. The resulting assignment matrix $S^{(l)} \in [0,1]^{N_l\times K_l}$ contains soft node-to-patch probabilities, with each row summing to one, rather than fixed binary memberships.

During the forward pass, node features are pooled using the column-normalized soft assignment matrix. Column normalization makes each pooled representation a weighted average of its contributing nodes rather than a sum, preventing large patches from producing larger feature magnitudes only because they receive contributions from more nodes. Coordinates are pooled in the same way to define the representation used at the next hierarchy level. Supplementary Figure 1 illustrates this clustering and centroid-aggregation step schematically.

$$X^{(l+1)} = ({S^{(l)}}_{norm})^{T} X^{(l)}$$

$$C^{(l+1)} = ({S^{(l)}}_{norm})^{T} C^{(l)}$$

During training, a small random perturbation is added to the finest-level distance matrix when candidate neighbours are constructed. This operation improves robustness to small changes in distance-based neighbour selection and does not reinitialize or replace the learnable pooling assignments.

## Causal Transformer and sparse dynamic graph learning

Each pooled or node-level temporal window is embedded and passed through a Causal Transformer encoder. Sinusoidal positional encodings are added to the temporal embeddings to preserve order within the look-back window. Multi-head self-attention is computed with a causal mask $M$, where $M_{ij} = -\infty$ when $j > i$ and 0 otherwise. This mask prevents the representation at a given time step from attending to future positions in the window.

$$PE(pos, 2i) = sin(\frac{pos}{10000^{\frac{2i}{d_{model}}}}),\ PE(pos, 2i+1) = cos(\frac{pos}{10000^{\frac{2i}{d_{model}}}})$$

$$Attention(Q, K, V) = softmax(\frac{QK^T}{\sqrt{d_k}} + M)V$$

The encoder uses a Pre-LayerNorm architecture for training stability. Layer normalization is applied before self-attention and before the feed-forward block, and residual connections are retained. For a hidden representation Z, a single encoder layer is computed as follows:

$$Z' = MSA(LN(Z)) + Z$$

$$Z_{out} = FFN(LN(Z')) + Z'$$

Sparse structural learning is implemented by basis decomposition. Instead of learning a separate dense $N \times N$ matrix for every feature channel, GeoDCD learns $N_{basis}$ shared adjacency bases $B_k$ and a channel-to-basis coefficient matrix $c$. The effective graph weight is reconstructed as a weighted combination of these bases, substantially reducing structural parameters while preserving multiple reusable interaction patterns.

$$W_{eff} = \sum_{k=1}^{N_{basis}} c_{\cdot,k} \otimes B_k$$

The latent state is propagated through the reconstructed graph layer and then mapped to a one-step prediction by a linear prediction head. Training minimizes a composite objective with a mean-squared prediction term and an L1 penalty on the learned adjacency. The prediction term implements the neural Granger criterion; the sparsity term discourages dense explanations that can fit temporal correlations without yielding interpretable causal structure.

$$Z_{next} = tanh(Z \cdot W_{eff})$$

$$L_{total} = \frac{1}{T}\sum_t \|x_{t+1} - \hat{x}_{t+1}\|^2 + \lambda_{L1} \|A_{learned}\|_1$$

## Input-output Jacobian sensitivity extraction

After training, GeoDCD extracts dynamic causal effects by automatic differentiation. For each inference window and target node i, the scalar or feature-valued prediction $\hat{x}_{i,t+1}$ is differentiated with respect to the source input $x_{j,t}$. The Frobenius norm of this derivative gives $S_{i\leftarrow j}(t)$, which is interpreted as the local sensitivity of the target future state to the source current state under the learned nonlinear dynamics. A non-zero value indicates that the fitted prediction of target node $i$ is locally sensitive to the observed history of source node $j$ at the evaluated state. We interpret this quantity as a state-dependent neural Granger sensitivity within the fitted predictive model.

The full Jacobian is computed by iterating over target nodes and retaining the gradients with respect to all source nodes. This produces a dynamic adjacency tensor with dimensions $N \times N \times T_{eval}$. For static graph evaluation, $S(t)$ is aggregated over time by averaging or by cumulative edge strength, then thresholded or ranked to compute binary edge metrics. For real-system visualization, time-resolved slices are retained to inspect regime changes such as ENSO phase differences, traffic peak transitions or EV-traffic coupling reversals.

This readout provides a directed neural Granger sensitivity graph for the fitted predictor under the observational assumptions stated below. It does not by itself establish interventional causality or remove the need to consider hidden confounding, measurement resolution and simultaneity. The Methods specify how the directed sensitivities are computed, whereas the Discussion defines the scope of their causal interpretation.

## Datasets and evaluation

Controlled benchmarks evaluate graph recovery when a reference topology is available. VAR provides a linear multivariate baseline with known coefficient matrices. Lorenz-96 tests nonlinear chaotic dynamics on a circular coupling manifold. Cluster-Lorenz introduces spatial patches with dense within-cluster and sparse between-cluster links, making it a direct test of whether the geometric prior helps recover multi-scale causal structure. Finance uses synthetic financial time series with random virtual coordinates to test robustness when geometry is deliberately uninformative.

Real-world evaluations test scale, interpretability and transfer. NCEP/NCAR sea-level pressure is processed on a 2.5°×2.5° global grid, yielding 10,512 nodes with latitude-longitude coordinates. Urban traffic experiments include San Diego, Beijing, San Francisco Bay Area and Greater Los Angeles networks. For the Shenzhen ST-EVCDP dataset, 18,061 public EV charging stations are aggregated into 247 traffic analysis zones; road occupancy and charging energy volume form a 494-node dual-modality graph using shared zone coordinates.

For controlled benchmarks, edge recovery is reported with AUROC, F1 score and Structural Hamming Distance. SHD counts the minimum number of edge additions, deletions or reversals required to transform the learned graph into the reference graph.

$$d_{SHD} = d_{add} + d_{del} + d_{rev}$$

For real-world systems without complete ground-truth graphs, evaluation combines scalability, predictive consistency and agreement with known physical mechanisms. Climate outputs are assessed against established teleconnection and circulation knowledge. Traffic and energy-mobility outputs are assessed by dynamic consistency with congestion regimes and spatially interpretable infrastructure zones. Forecasting error is used only as a supporting diagnostic: the primary object of analysis is the inferred causal structure.

### Baseline methods and implementation

GeoDCD is compared with linear and nonlinear causal discovery baselines under matched preprocessing and evaluation protocols. Baselines include GVAR, cMLP, cLSTM, TCDF, eSRU, NAVAR variants with MLP and LSTM encoders, CUTS+, UnCLE and Causalformer. Synthetic benchmark comparisons use the same ground-truth adjacency matrices and random-seed protocol across models. Real-world experiments preserve chronological train/validation/test splits to prevent temporal leakage.

The implementation uses PyTorch. The Causal Transformer encoder uses dmodel = 128, four attention heads and two layers. Basis decomposition uses $N_{basis} = 4$. Models are trained with Adam using a learning rate of 3●10−4, cosine annealing over 100 epochs and L1 sparsity weight λ L1 = 0.01. Each quantitative experiment is repeated with five independent random seeds, and all reported metric values are means across seeds.

### Causal assumptions

GeoDCD estimates directed predictive dependencies under temporal-ordering, observed-conditioning and model-adequacy assumptions. A link indicates that the fitted prediction changes with a source variable after conditioning on the measured multivariate context. It should be interpreted as a neural Granger edge within the observed system, not as an interventional effect guaranteed under arbitrary hidden confounding.

Hidden common drivers, missing variables, temporal aggregation, synchrony and modality-specific noise can affect graph recovery. The method is strongest when relevant drivers are observed at sufficient temporal resolution and when the coordinate embedding is meaningful for the system. These assumptions are stated explicitly so that synthetic benchmarks, climate analyses and energy-mobility results are interpreted under the same causal-discovery frame.

## Data availability

All datasets used in this study are accessible as follows:

- **Synthetic data.** The VAR, Lorenz96, and Cluster Lorenz datasets are generated using the synthetic data generation scripts provided in the GeoDCD-core repository at https://github.com/BUAA-ACME/GeoDCD-core. The Finance dataset is generated using the UnCLE repository at https://github.com/etigerstudio/uncle-causal-discovery.
- **Meteorological data.** The ncep-slp dataset comprises daily mean sea-level pressure fields from the NCEP/NCAR Reanalysis 1 project, provided by the NOAA Physical Sciences Laboratory at https://psl.noaa.gov/data/gridded/data.ncep.reanalysis.html.
- **Traffic data.** The San Diego, Greater Los Angeles, and San Francisco Bay Area traffic datasets are subsets of the LargeST benchmark, publicly available at https://github.com/zezhishao/LargeST. The Beijing traffic dataset was provided by Amap (Gaode Maps) and cannot be redistributed due to commercial data licensing restrictions. The ST-EVCDP dataset is an open-source electric vehicle charging demand dataset, available at https://github.com/IntelligentSystemsLab/ST-EVCDP.

# Acknowledgements

The research is funded by the National Natural Science Foundation of China (No. T2588101) .

# Author contributions

Y.H.Y. and K.Q.Z. designed the study, analysed the data and wrote the first draft of the paper. X.L.M. designed the study, supervised all parts of the analysis, analysed the data and edited the paper. Y.P.W. and W.P.W. edited the paper. All authors read and commented on the final manuscript.

# Competing interests

The authors declare no competing interests.